\documentclass[a4paper]{article}
\usepackage{graphicx, color}
\usepackage[utf8]{inputenc}
\title{Open Research Issues and Tools for Visualization and Big Data Analytics}
\author{Rania Mkhinini Gahar$^1$,$^2$  \and Olfa Arfaoui$^1$,$^3$  \and Minyar Sassi Hidri$^4$}

\date{
	$^1$National Engineering School of Tunis, University of Tunis El Manar, Tunisia \\
$^2$OASIS Research Laboratory\\
$^3$RISC Research Laboratory\\
	$^4$Computer Department, Deanship of Preparatory Year and Supporting Studies, Imam Abdulrahman Bin Faisal University, Dammam,
Saudi Arabia\\ %
}
\begin{document}
\maketitle
\begin{abstract} The new age of digital growth has marked all fields. This technological evolution has impacted data flows which have  witnessed a rapid expansion over the last decade that makes the data traditional processing unable to catch up with the rapid flow of massive data. In this context, the implementation of a big data analytics system becomes crucial to make big data more relevant and valuable. Therefore, with these new opportunities appear new issues of processing very high data volumes requiring companies to look for  big data-specialized solutions. These solutions are based on techniques to process these masses of information to facilitate decision-making. Among these solutions, we find data visualization which makes big data more intelligible allowing accurate illustrations that have become accessible to all. This paper examines the big data visualization project based on its characteristics, benefits, challenges and issues. The project, also, resulted in the provision of tools surging for beginners as well as well as experienced users.
 \vspace{1mm}
\end{abstract}
\section{Introduction}
\label{sec:introduction_and_overview}
Every day we generate 2.5 trillion bytes of data. In fact, 90\% of the world's data was created in the past two recent years only \cite{torphy2020social}. The data comes from different sources such as the sensors used to collect climate information,
social Media Posts, digital images, and online videos,
online Purchase, transnational records and GPS (Global Positioning System) signals from mobile phones. This data, which has resulted essentially from the meeting of three elements which are Internet, social networks, and smart devices (computers, tablets, smartphones, connected objects...), is called big data or massive data. It is considered very interesting according to the pertinent information that may contain. Actually, we note that 1\/3 of business leaders' decisions are based on information they don't trust or don't have and half of them say they don't have access to the information they need to do their job and 83\% of CIOs (Chief Information Officer) cite analytical business intelligence (BI) as part of their plans to improve their competitiveness.

Moreover, 60\% of CEOs (Chief Executive Officers) need to improve the capture and understanding of information in order to make decisions more faster \cite{walter2021indigenous}. For example, data could be analyzed to i) detect customer feelings and reactions or critical or life-threatening conditions in hospitals to intervene in time; ii) predict weather patterns to plan the optimal use of wind turbines and make decisions based on transactional data in real-time;  iii) identify criminals and threats from videos, sounds and data streams; iv) studying student reactions during a lesson and predict which ones will succeed in the basis of statistics and models gathered over the years (big data in the education field).

Research on big data analytics is entering a new phase called fast data where multiple gigabytes of data arrive in the big data systems every second \cite{08333,HidriAH22,AlsaifHH21,AlsaHH22,ZoghlamiHA16,GaharAHH18}. Modern big data systems
collect inherently complex data streams due to the 3 basic Vs which are Volume, Velocity, and Variety and to which are added Veracity, Validity, Vulnerability, Volatility,
Visualization which consequently give rise to the 10Vs \cite{manogaran2017big} of big data. The well-designed big data systems must be able to deal with all 10Vs effectively by creating a balance between data processing objectives and cost (i.e., computational, financial, and programming efforts). Data collection and storage capabilities have allowed researchers in diverse domains to collect and observe a huge amount of data. However, large data sets present substantial challenges to existing data analysis tools \cite{gahar2019distributed}.

We will focus in our paper on one of the most important big data' Vs which is data visualization.

One of the most crucial and useful tools for comprehending corporate information is data visualization. Because of this, a picture truly is worth a thousand words. Data has been visually represented by humans for hundreds of years. We've collected data, organized it, and presented it in maps, charts, and graphs to tell a richer and deeper story than it may have otherwise. The data boom coincides with the technological boom. Additionally, we have been able to process ever-increasing volumes of data quickly thanks to the same technology. Although they might not be immediately apparent in the first text format, trends, patterns, and other insights are quickly identified utilizing data visualization software.

The most effective strategy changes to visual data displays after reports and dashboards take their place since they can fit a lot of information into a little amount of space. Examining the extensive data sets and graphic presentations that enable quick and accurate translation might take hours, days, or even weeks. Thanks to advanced technology, many data visualization tools allow for interactive functions. This flexibility provides the ability to switch and change quickly, which helps the user to discover and learn about alternative viewpoints. This comprehensive, interactive presentation can rarely be achieved quickly by processing raw data without visualization software. The information industry frequently faces the difficulty of the quantitative component.

Knowing that decisions are made as a result of visual representations requires a solid comprehension of data. In the absence of context, visuals are ineffective. But there is an easy target is just to let the workers and the tools perform their jobs. As long as you utilize the appropriate tools and the individuals conducting the data analysis are aware of where the data originated from, who can use it, and how it will be used. The data visualization will next be translated, processed, and on a more clearer course for making those important decisions. Data visualization's significance in the world of corporate information is being realized more and more every day. They have the ability to not only supply useful data but also understand how to process it, which guarantees the organization stays competitive. This is because they are high-performance analytics tools that offer better ways to analyze data faster than ever before.

Visualization is critical in today's world.
Big data is difficult to visualize. Due to in-memory technology limitations and low scalability (scaling up), functionalities and development time response, visualization tools, and current big data are faced with technical challenges. Traditional graphs cannot be relied upon to attempt to plot a billion data points.
We, therefore, need different ways of representing data. If we take into account the multitude of variables resulting from the
variety and speed of big data and the complex relationships that
unite, we can see that developing a visualization
significant is not so easy.

Spreadsheets and reports stuffed full of numbers and algorithms are far less successful at communicating meaning than reports and charts that show enormous amounts of complex data. Because of the constraints of in-memory technology and their inadequate scalability, functionality, and response time, current large data visualization solutions suffer technological difficulties. When plotting a billion data points, we can't rely on typical graphs, thus we need alternative methods, such as data clustering or the use of treemaps, sunbursts, parallel coordinates, circular network diagrams, or cone trees. \cite{khalid2021big}.

The rest of this paper is organized as follows: The DataViz' presentations are presented in section 2. Section 3 presents
its benefits. Section 4 introduces the characteristics of such DataViz project. Section 5 discusses some DataViz tools surging for beginners as well as well as experienced users. Section 6 revealed the main data visualization challenges. The DataViz issues are highlighting in section 7.
Overall discussion with limitations are stated in section 8.

\setlength{\parindent}{5mm}

\label{sec:background}
\section{The DataViz and you: presentations}
Many data scientists define data visualization in different ways. Indeed, they agree that it is indeed a visual form to visualize by facilitating access to it. Another data experts' group agrees that data visualization is meaningless if it does not encompass understanding, exploitation and decision-making, speed, and information sharing.

\subsection{DataViz features}
The value of data visualization lies in its ability to meet three main imperatives namely \textit{interpretable}, \textit{relevant} and \textit{innovative}.

\begin{itemize}
    \item \textit{Be interpretable}: In a context where the volume of data is exploding with the exponential growth in the use of the Internet and in particular
Google, so-called \textit{unstructured} data experiences the same
evolution. But a data visualization starting from these data which would not be interpretable, that is to say clear, would be useless. There must be some clarity regardless of the volume or source of the data \cite{lensen2020genetic}.
\item \textit{Be relevant}: At a time when big data is a central issue for companies, several techniques to process this mass of information in a relevant way must be put in place.

Relevance is linked to interpretability. The data visualization must make it possible to answer questions in a defined context and aimed at specific objectives. Data sources must be reliable. Data integrity is the basis for meaningful data visualization. You must ensure that your information is correct and up to date. It is necessary to sort the data for optimal analysis and to consider using all the data at your disposal. This allows you to cross-reference information and thus bring out a more complete analysis to better support your digital marketing department\cite{skender2023investigation}.

\item \textit{Be innovative}: Finally, data visualization is only of interest if it brings
new information, and originality, if it gives a perspective
unpublished on a subject. It provides a different perspective, to illustrate an analysis \cite{ibeh2023dataviz}.

\end{itemize}

\subsection{Brief history of DataViz}
Originally, it was a simple human limitation that spawned the Data Visualization approach: our brain is simply unable to easily process large volumes of raw data to extract useful information. Maybe we can do it occasionally, but certainly not every time, let alone many times every day.

However, Data Visualization is not just about the graphical representation of data. It is also a story that is told with the help of these representations.

The first successful expression of this approach is a flow map that tells the story of the Russian campaign led by Napoleon. We owe it to the engineer Charles Joseph Minard who represents, at the beginning of the 19th century, the story of the colossal human losses of the Russian campaign during which the Napoleonic army arrived in Moscow with less than a quarter of its starting squad.

Shortly after, nurse Florence Nightingale had the idea of using graphic representation to allow her reader to compare facts with complex correlations \cite{ibeh2023dataviz}.

\begin{figure}[ht!]
\centerline{\includegraphics[width=8.2cm]{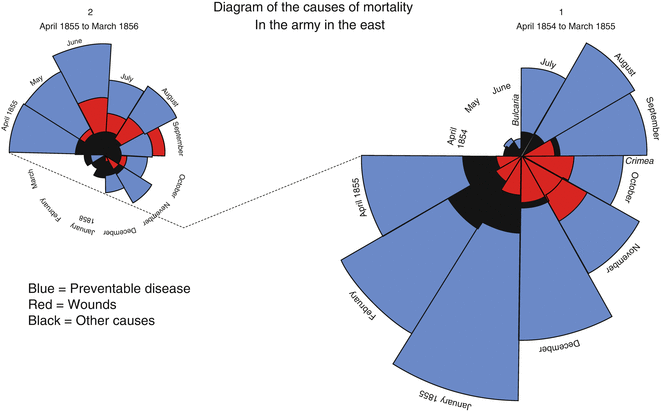}}
\caption{Florence Nightingale's Diagram \cite{friendly2008golden}.}\label{fig:florence}
\end{figure}

She presented, for the attention of Queen Victoria, the main causes of death of British soldiers engaged in the Crimean War. His graphic support, dated 1858, allowed him to eloquently highlight that epidemics were much more devastating on the workforce than the injuries suffered in combat.
Florence Nightingale has therefore used a graphic support of data, presented in figure \ref{fig:florence}, to communicate information, of course, but also to convince, that is to say, to orient the conclusion that one draws from it.

Until the 19th century, Data Visualization therefore evolved through isolated, independent attempts, which explored both the possible fields of application and the possible graphical approaches. In other words, the discipline was forging a vocabulary but it still lacked common rules that is to say, a grammar.

This grammar was laid down by Jacques Bertin who in 1967 developed the real bases of graphic language.

\begin{figure}[ht!]
\centerline{\includegraphics[width=8.5cm]{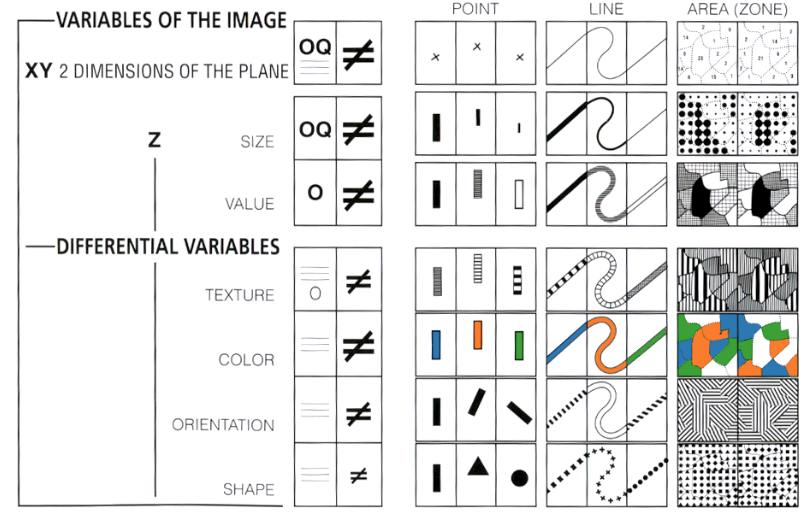}}
\caption{Jacques Bertin's graphic semiology \cite{morita2019reflection}.}\label{fig:graph}
\end{figure}

In figure \ref{fig:graph}, Jacques Bertin defined graphic semiology, i.e. the elements that can be modified in a Data Visualization to represent information. Identifying and clearly defining these graphic variables (color, size, surface) was simply the grammar that was missing from the graphic language.

It is by applying these new - but common - rules that we have gradually managed to remove superfluous graphic elements and define what characterizes a relevant data visualization.


\subsection{The DataViz is pretty, but what is it for?}
Data visualization concerns a wide range of business sectors. If we stick to our panel, several areas are represented, namely consumer goods, business services, industry, media, marketing and advertising, scientific research, public service, telecom, transport, logistics, etc... All companies use data visualization either to do
certain things they were already doing better (\textit{optimization}), or
to enrich their activity with value-added tasks (\textit{innovation}). In terms of optimization, DataViz allows, for example, to
\textit{Lagardere Active} to accelerate the production of its reports and \textit{STMicroelectronics} to make its manufacturing process more
efficient;

In the area of innovation, PagesJaunes discovers and rectifies, thanks to data visualization, the shortcomings of its indexing, \textit{Voyages-sncf.com} launches a new innovative service (Mytripset), \textit{Alcatel Lucent} imagines the mobile applications of tomorrow, etc.

\subsection{Dual purpose of optimization/innovation}
More specifically, 5 use cases illustrating this dual purpose
optimization and innovation: the use of DataViz to:
\begin{itemize}
    \item \textit{Improve the company management}: One of the first uses of data visualization is to contribute to
more effective management of activity and performance, oriented
towards action.

In our modern economy, all directions need to manage their organizations as finely as possible by relying on numerous and rich data and by producing relevant dashboards.

\textit{"A picture is worth a thousand words"} and nowadays \textit{"the image takes precedence often on words"} \cite{organisciak2022giving}.

Data visualization is the art of telling figures in \textit{creative} and \textit{gameful} way
 \cite{jung2021designing}.
    \item \textit{Improve the customer relationship}:  Marketing and customer relationship management are two functions of choice for data visualization. We visualize customer data to improve Customer relationship management (CRM) multi-channel.
In addition, DataViz is better to qualify the customer base.

    \item \textit{Define the company's offer}:  At the opposite end of the spectrum, data visualization provides companies with tools to better define their offers. Moreover,
the exploration of collected customer data and the ability to test different hypotheses prove particularly valuable.
    \item \textit{Contribute directly to the company business}:  The DataViz serves to better understand your competitive positioning. That's why Data visualization could be considered a relevant source of value and differentiation which immediately identify customer postponements. Here,  data visualization could directly influence the business economics model \cite{zhang2021interactive}.
    \item \textit{Empowering citizens}: Data visualization also finds its usefulness outside the walls of
the company. It can contribute to better informing the citizen and therefore giving him the means to act.

In fact, company can allow citizens to think through \textit{Citizen DataViz} projects may take a longer turn activist \cite{naerland2021towards, raineri2021innovation}. This is the case, for example, of the \textit{pariteur} of France Televisions.
\end{itemize}

\subsection{DataViz and Infographics}
The Dataviz takes the information a person needs and presents it in an easily understandable way. Infographics are a mix of Dataviz, journalism, and marketing. They use strategically chosen data visualizations and lexicons to explain a complex story easily. The confusion in terminology is understandable however the terms are not interchangeable. Both turn data into easy-to-understand visualizations. These tools are extremely powerful when it comes to explaining numbers in an educational way to people who are reluctant to analyze data. This is their only common point. Here is a definition for both forms of presentations. Some distinction points between these two terms can take place in Table \ref{tab:versus}.
\begin{table}[!ht]
\begin{center}
\scriptsize{
    \centering
    \caption{Infographics versus DataViz}
    \label{tab:versus}
    \begin{tabular}{p{5.5cm}p{5.5cm}}
    \hline
        ~~~~~~~~~~~~~~~~~~~~~~~~\textbf{Infographics} & ~~~~~~~~~~~~~~~~~~~~\textbf{DataViz} \\
        \hline\hline

             Promotes the information understanding that we already know by representing it in graphic form.&Bring out information that was unknown by analyzing data presented in graphical form.\\
            Modest data amount.&Huge data volume.\\
              Good design makes a product useful.&Good DataViz makes information useful.\\
            Good design is aesthetic.&Good Data is aesthetic.\\
            Good design makes a product understandable.&Good DataViz makes information understandable.\\
             Good design is honest.&Good DataViz is honest.\\
            A didactic approach focused on others.&Self-knowledge tool.\\
            Understanding support.& Decision support.\\
        \hline
    \end{tabular}
    }
    \end{center}
\end{table}

\subsection{Main reasons for using DataViz}
Three main reasons explain the use of data visualization namely
\textit{confirm or refuse hypotheses on a market}, \textit{educate} and \textit{explore}.
\begin{itemize}
    \item \textit{Confirm or refuse hypotheses on a market:} The DataViz can then take the form of a dashboard, making it possible to decide while having a global vision of the studied market.
    \item \textit{Educate:} Internally, companies use DataViz for research work reporting or brainstorming sessions. It can be a good complement to creative approaches such as \textit{gamification}.
    \item \textit{Explore:} This is the most futuristic aspect of data visualization, which certainly will develop. Dataviz can help build predictive models. We are then in the field of data analysis
\end{itemize}

\subsection{DataViz: buzzword or real innovation?}
The data that are thus available to professionals to guide them in their decisions are increasingly numerous and multi-structured. But how not to be overwhelmed and make it a real tool for reflection and decision-making? How to obtain answers to fundamental questions whose answers are for the moment unknown?

It is in the face of these requirements that DataViz takes on its full meaning. The representation of data in the form of images makes it easier to understand them. There are several definitions of DataViz in the academic and industrial state of the art. These different definitions all converge on the fact that DataViz is a way to give meaning to data in order to extract information from it and therefore exploit it. Dataviz not only enables intellectual understanding but also transforms a set of raw data into actionable information.

In addition, it accelerates the understanding, decision, and action that we have just mentioned. It is also a mode of communication, allowing data not to remain confined to the world of BI or statistics but to infuse the entire organization and become a support for decision-making and collaborative work.

DataViz has many uses and leads to a variety of benefits for the organization. First, it contributes to more effective management of activity and performance, oriented towards action. This improvement in management is manifested by taking a step back in addition to other tools whose horizon is in the shorter term. In other words, DataViz can be used as a decision-making, and strategic tool, usable by a local manager to manage his performance.

Another important use of DataViz is the reinvention of customer service to improve its efficiency. To improve the management and understanding of their KPIs (Key Performance Indicators), SFR uses DataViz to identify causal relationships in their data sources in order to find hidden patterns through their main sales channels. A good customer relationship is based on perfect knowledge of the customer himself. What are their characteristics and behaviors? How to segment and classify them? The exploration capabilities in the data enabled by data visualization find their full meaning in providing answers to these questions.

Another key point about DataViz is its ability to foster innovation and its potential to get the business to consider new possibilities. In particular, it is a testing ground for new modes of interaction with users.

\section{DataViz's benefits}
In a context of ever-increasing and often highly complex volumes of data, DataViz has many advantages. Data visualization is far from being an accessory intended to embellish your website or your presentations. Synthetically, we can say that DataViz improves the data understanding, the data communication, the decision-making, and the ability to innovate.



\subsection{Dataviz makes it easier to understand data}
With big data advent and the proliferation of data sources, companies are increasingly using data visualization. These visual representations make it easier to understand raw data and thus help in decision-making. Big data is not merely \textit{more data}; it is data that is so vast, so varied, and collecting so quickly that typical procedures and methodologies, including "normal" software like Excel, Crystal reports, or other programs, are ineffective. DataViz makes it possible to make the most comprehensible data important and what they mean, regardless of the audience
concerned. Its effectiveness is based on the fact that a majority of us grasp and retain information better when it is represented visually. The following image illustrates this fact, which was studied by an American psychologist.

Unlike a table filled with figures, DataViz helps to highlight information that seems complex or drowned in a large number of parameters. The following example illustrates this fact well. For example, we want to analyze the life expectancy by country. Table \ref{tbl:visits} presents an extract of values from the top 5 countries since the values file is very large.
\begin{table}[!ht]
\centering
\caption{Life expectancy per country.}\label{tbl:visits}
\scriptsize{
\begin{tabular}{llllll}
\hline
& Country & Life expectancy & ISO-code \\
\hline\hline
0 & Afghanistan   & 64.5&AFG   \\
1 & Algeria   &76.7  &DZA  \\
2 &Andorra   & 81.8&AND   \\
3 & Angola   & 60.8&AGO   \\
4  & Antigua and Barbuda  & 76.9&ATG  \\
\hline
\end{tabular}
}
\end{table}

But is it ideal for ordering easily and
between countries quickly? And explain it in a few
seconds to his audience? If we transcribe this information on a map, with more or less bright colors depending on the strength of the index, everything becomes clearer. We want to quickly understand which countries have the highest rate of life expectancy.

\begin{figure}[ht!]
\centerline{\includegraphics[scale=0.6]{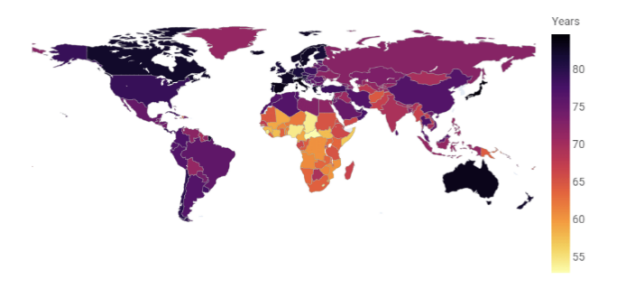}}
\caption{Life expectation by country (Source: Kaggle / WORLD DATA by country (2020).}\label{fig:visits1}
\end{figure}

Figure \ref{fig:visits1} allows us to assess at a glance countries with the highest rate of life expectancy. Analyzing an Excel file is much faster by visualization, especially if the data is large and complex.

Our brain needs less than 250 milliseconds to
enter (1), understand (2), and respond (3) to information under
visual form. Whereas comparing raw data in tabular form
requires an effort of memory that quickly reaches its limits. Ultimately, data visualization invites us to take up the classic distinction between data, information, and knowledge.
If the data are unitary, raw elements, reflecting reality, the information is their coherence to give them meaning.

\subsection{Dataviz improves communication}
Data not only reflect reality, but they also are not just a steering lever. They are also communication tools. Unfortunately, few of us are fluent in our language. Most of us need an interpreter to make us penetrate the intelligence of the data.

This is where data visualization comes in. Moreover, Communication is the major asset of the DataViz for
coordinate the teams. Indeed, the infinite volumes of data stored by companies are generally only within the reach of data analysts and other technical profiles. But once put into images, this data is accessible to everyone without any prior training. The support generated by DataViz makes it possible to unify the discourse and convey unambiguous messages.
  We can truly speak of democratization of access to
given \cite{holbrook2019open}.

\subsection{Dataviz optimizes and accelerates decision-making}
This is the logical continuation of an easier understanding of the data. DataViz allows the development of interactive dashboards. Unlike static charts, such as those in Excel,
interactive DataViz allows the exploration of data in
depth, with less effort. In a few clicks, it is now possible to release correlations between the operational actions put in place, the performance, and their impact.

It is also much easier to compare its indicators to the market, and competitors. Once the data has been clarified and better identified, decision-makers can focus on the essentials and make choices in a way more simple.

\subsection{Dataviz promotes innovation}
Finally, we cannot close this part on the benefits of data
visualization without examining its potential in terms of innovation. Indeed, DataViz is also a field of research that can
encourage the company to consider new possibilities. In particular, it is a testing ground for new ways of interacting with users \cite{sarica2019data}.
\section{Characteristics of DataViz projects}
The deployment of DataViz software is a crucial step that must be perfectly orchestrated to guarantee the success of the project. This later must be both fast and light \cite{burnett2021destabilising}.

\subsection{DataViz project's speed}
The first characteristic of DataViz projects is that they are,
all in all, quick to conduct. Their big advantage is in particular to reduce the time between the launch of the project and the ability to show a first operational version \cite{raineri2021innovation}.

There are several reasons for this speed. One of them is that DataViz projects require hardware and software resources that interfere little with existing architectures. They, therefore, do not require long and complex budget validation cycles: the necessary environment can be available in a short time. Another reason is that data visualization lends itself well to
POC (Proof Of Concept) type approaches experimentation, trial and error iteration loops. Dataviz projects have a very empirical side, completely in phase
with current development approaches, such as agile methodology, rapid Agile development, or Scrum.

\subsection{DataViz project's lightness}
Lightness is involved in the technical resources required which
are quite light. Example: Recent technological developments, such as the development of JSON-type formats (DS.JS3), put us in a direct connection with data. Thanks to these formats, we can recover varied data from all horizons, using standard applications. A real human-data interface is thus being established.
\subsection{Success key factors}
To succeed in a data visualization project, it is necessary to bring together key success factors that can be classified into three categories. First of all, there are the classic good practices of any project: ensuring preparation and planning, choosing the right scope, implementing the appropriate methodologies, etc. The second category is that the DataViz project concerns data: their targeting, quality, respect for confidentiality, and access authorizations, are therefore essential. Finally, of course, ergonomics and graphic intelligence play a
key role in the acceptance of DataViz and its effective use (although, by the way, these aspects should be part of any IT project, etc.).

\subsubsection{Prepare the project well}
 The first secret to success is preparation. This is understood on two levels: the content of the data visualization, on the one hand, and the project approach, on the other hand.

 DataViz is not a panacea and, to be useful, it must be well thought out. On the methodological aspect, the preparation consists of implementing establishes a process for collecting, analyzing, and representing data that is viable over time, but also sufficiently flexible.
\subsubsection{Target data and visualizations according to the profile of the
users}
A second success factor is related to the nature of the data visualization, namely a communication tool. However, for communicating effectively with someone, it is important to meet his need, with clarity and in a form he understands.

Three key questions to ask yourself to choose the best
representation:
\begin{itemize}
    \item \textit{What question do we want to answer?} All the graphs do not make it possible to present the same analysis (distribution, evolution, decomposition...). Hence the importance for the designer to question his intention.
    \item \textit{Who are we talking to?} Is he an expert or a layman? What should he do with the information (e.g. retain the information for later or make an immediate decision)?
    \item \textit{In what context is the interlocutor?} The good reception of the graphic does not only depend on the
graphic itself, but also the intellectual and visual availability of the reader. All that the graphic designer can do is try to anticipate this greater or lesser availability, in order to choose the most suitable representation.
\end{itemize}

The difficulty increases when the data visualization must address different audiences. It is then necessary to provide modes of representation adapted to each of them. This is the case, for example, of the Belgian FPS Economy, which communicates with both the general public and professionals.

\subsubsection{Start on small perimeters, to learn}
Another good practice is to "get your hands dirty" on first
restricted perimeters, if possible as controlled as possible. This allows you to move forward, without too many risks, in trial and error mode until a first satisfactory solution is obtained.

\subsubsection{Ensure data quality at source}
In BI, you reap what you sow or, to put it more lapidary way: garbage in, garbage out. In other words, if we want data visualization to be able to communicate the right messages, make it possible to make informed decisions, to explore unknown territories, there is one condition to be met above all: to have quality data entrance.

In return, DataViz improves the quality of the data. First, because it compels a certain discipline; then, because it also visualizes... the non-quality of the data. Repeated outliers may appear at first glance as oddly placed dots, for example.

\subsubsection{Focus on cooperation between several departments}
Another ingredient of success lies in the cooperation between
the actors of the project. DataViz thus contributes to breaking down the silos that may exist in the company and contributes to greater cross-functionality.

\subsubsection{Train the teams}
Data visualization does not require side training
users. On the contrary, we can say that it is fully successful
when it is immediately adopted. For this, \textit{simplicity} and \textit{intuitiveness} are essential. But offering a simple rendering can be extremely complicated. This is why the training of those who produce the visualizations is an \textit{undeniable} plus.
\subsubsection{Using aesthetics as a lever for appropriating information}
Data visualization cannot be reduced to a representation
aesthetics of data. One can make pretty representations that are perfectly useless. But that's not to say that aesthetics don't play a role. Used well, it is an essential criterion of efficiency for DataViz. In this concern to combine aesthetics and efficiency, companies have every interest in being imaginative and going beyond traditional Excel-type charts, if that makes sense.

\subsection{Pitfalls to avoid}
If certain good practices maximize the chances of succeeding in your data visualization project, you can expect, as with any project, to encounter difficulties such as:
\begin{itemize}
    \item  The risk of overloading it with information.
    \item  A general management that does not necessarily perceive an interest of data visualization immediately.
    \item  Skepticism about the performance of the tool.
    \item  A pitfall to avoid: forget your classics: We should not confuse simplicity with simplism \cite{deng2005building}.
\end{itemize}

\subsection{The DataViz's impact on the relationship between IT and business lines}
Data visualization projects have the particularity, as we saw previously, of offering great autonomy to users.
users. This is why, in this area, the relationship between IT
and the Professions are set to evolve. It has happened that friction has arisen, with the IT Department feeling deprived of some of its prerogatives. These tensions have unfortunately been maintained by some providers of DataViz solutions by addressing only the Business Lines without going through the IT Department to win contracts \cite{grant2019pretty}.

Thus, we cannot speak of a loss of prerogatives of the DSI (Dimensional Strategies Inc.) or the BI teams. Simply, data visualization raises new questions about how to represent data, about the distribution of roles and responsibilities between the business lines and the IT department, and about how to conduct projects \cite{graessley2019industrial}.

CIOs understand this. Even those who were initially reluctant are realizing that data visualization is not a threat, and are softening their stance
\cite{santolalla2020dataviz}. In truth, data visualization is a chance for CIOs and
BI teams. On the one hand, it will relieve them of time-consuming tasks, allowing them to focus on their missions with higher added value. On the other hand, they even have a unique opportunity to invent a new form of BI and relationship with business. What we can remember is that data visualization, even if it
provides great autonomy to the business lines, is a question that should interest the IT department and the BI teams, quite simply because it touches the data. Autonomy of businesses is useful if it is implemented smartly and if it allows them to obtain even more value from the CIO and the BI. Presumably, by accustoming the Professions to speak the language of data, thanks to graphics, the DataViz will contribute to the taking of awareness of the value of data. It can therefore play a role unifier, at the service of business creation.
\section{Which tools for which data visualization?}
Information visualization faces increasing hurdles as the big data era progresses. First of all, the amount of data that needs to be visualized surpasses the size of the screen. Second, a typical computer cannot be used to store and process the data.

A big data visualization solution needs to offer perceptual and performance scalability to solve both of these issues.

In this section, we will focus on some data visualization tools for beginners as well as for experienced users.

\subsection{Tools accessible to beginners}
Tools for beginners are available to allow them to create DataViz without resorting to programming or its basis and no expertise is required.

\subsubsection{Office software and extensions}
\paragraph{\textbf{Excel}} Excel remains one of the basic tools for data visualization \cite{oike2019simple,patel2022data}. The maximum number of values in a column is about 1,999,999,997  \cite{hiljazi2018developing}.

Third parties create Excel add-ins to provide Excel users with extended functionality and save their time and effort. Developing these add-ins requires coding expertise in languages such as XML (eXtensible Markup Language) and VBA (Visual Basic for Applications) and providing an easy-to-use interface that complements Excel.

Table \ref{tbl:excel} overviews some Excel add-ins for DataViz.
Table \ref{tbl:excel} overviews some Excel add-ins for DataViz.

\begin{table}[!ht]
  \centering
  \scriptsize{
   \caption{\label{tbl:excel}Some Excel add-ins for DataViz.}
  \begin{tabular}{p{3.5cm}p{8cm}}
    \hline
   \textbf{Add-in} & \textbf{Description}\\ \hline\hline
   Filled Map&Used to display high-level chart data within a map.\\
{3D-Mapping}&A three-dimensional (3D) data visualization tool called Microsoft 3D Maps for Excel enables you to see the data in novel ways. \\
    Bing Maps& With the Bing Maps add-in, Excel users can quickly plot locations and display their data using Bing Maps.\\
Radial Bar Chart&It displays data from standard bar charts on a circular pattern. With this representation, visual analysis gains the advantages of both bar charts and circle graphs. The Radial Bar Chart has a distinctive style that makes it a very adaptable representation.\\
XLMiner Data Visualization App&It instantly visualizes data in your Excel spreadsheet and lets you adjust the variables plotted on each axis, zoom in and out, add filters to highlight important data, and use textit size by and textit color by categorical variables for quick insights. \\
\hline
  \end{tabular}}
\end{table}

\paragraph{ \textbf{LibreOffice} }LibreOffice is a free and open-source office suite, derived from the \textit{OpenOffice.org} project, created and managed by \textit{The Document Foundation}. Extensions are provided to enable various activities, including visualization. LibreOffice extensions are software add-ons that you can install in addition to the core LibreOffice apps to extend the capability of the suite in one or all of the programs (Writer, Calc, Impress, etc.).



Some examples of LibreOffice extensions dedicated to data visualization can be presented in the table \ref{tbl:libreoffice}.

 \begin{table}[!ht]
  \centering
  \scriptsize{
   \caption{\label{tbl:libreoffice}LibreOffice Extensions.}
  \begin{tabular}{p{3.5cm}p{8cm}}
    \hline
    \bf Extension &  \textbf{Description} \\ \hline\hline
  GeoMap&Inserts map images directly into your document from address information.\\
 GeOOo& His free solution suggests using the LibreOffice tools to create thematic maps.\\
 ClusterRows&A  Is a LibreOffice Calc extension that groups rows into clusters and colors the clusters to show them.\\
 OOoHG Gallery&Includes a \textit{Gallery} that contains 1600 maps, diagrams, and graphs for geography and history organized into 96 categories (in bitmap and vector graphic format).  \\
OpenStreetMap Presentation&It uses penStreetMap Datas as background for nice presentations.\\
\hline
  \end{tabular}}
\end{table}

\subsubsection{Online office suites}
 When choosing between data visualization tools, one option worth considering is Google Sheets. Google's spreadsheet application can be used to generate charts, tables, and even maps that can be embedded in a website. They're easy to make and can be configured to update automatically \cite{dougherty2021hands}.

It's not for every visualization need. There are some tasks that need for more intricate data visualization strategies and customisation than what Google Sheets can offer.


Google drive extensions can also help novice users to perform data visualizations, namely Fusion Tables \cite{doshi2014review}, Slemma \cite{cardona2017evaluacion}, Geckoboard \cite{orlovskyi2019using}, VizyDrop \cite{minelli2013big}, BIME Analytics \cite{ur2016big}, Cyfe \cite{oliver2020kiserleti}, and Datahero \cite{atwood2018using}, etc.

\subsubsection{Office 365}
Microsoft 365 is made up of the Office suite (Word, Excel, PowerPoint, Outlook, OneNote, Publisher, and Access), as well as a set of online services such as OneDrive, Exchange Online, SharePoint Online, Teams, and Yammer. The Office suite allows work in offline mode like a perpetual suite, which distinguishes it from Office Online, which is used from a Web browser. The principle of Microsoft 365 is to be updated as new versions of Office are released \cite{wilson2014microsoft}. It provides also some integration apps to visualize your data in an interpretable, innovative and relevant way. Table \ref{tbl:365} describes someones highlighting their main functionalities.

 \begin{table}[!ht]
 \begin{center}
  \scriptsize{
   \caption{\label{tbl:365}Data Visualization Apps Integrated with Microsoft 365.}
  \begin{tabular}{p{2.8cm}p{8.5cm}}

    \hline
    \bf Apps & \textbf{Description}  \\
    \hline\hline
  Workday Adaptive Planning &
     Is a cloud-based enterprise planning platform that provides modeling, budgeting, and forecasting capabilities to enable firms collaborate on planning. It offers capabilities like workforce modeling, balance sheets, and spending management, among others.\\ \\

Coras     &
 It is an enterprise decision management platform suitable for IT, government, legal, and marketing teams. It can be adapted to any type of business and easily mapped to existing processes and governance.\\\\

 Microsoft Power BI&
It enables the creation of personalized and interactive data visualizations with an interface simple enough for end users to create their reports and dashboards.\\\\
Cyfe     &
Users can monitor and analyze data spread across all of your web services, including Google Analytics, Salesforce, Google Ads, MailChimp, Facebook, Twitter, and more, from a single spot in real-time with this all-in-one dashboard software.
       \\ \\
DBxtra  &
It is an ad-hoc reporting and business intelligence system that gives companies the tools to create and distribute unique reports on key performance indicators. It has capabilities including an Excel reporting service, a report and dashboard designer, online report distribution, and a report scheduler.

        \\ \\
  MicroStrategy Analytics
     &
 It equips businesspeople with self-service tools to study data and share insights in minutes, enabling them to make quicker, smarter business decisions.
       \\ \\
     Kepion
     &
It is a Microsoft BI-powered cloud-based business planning tool. In a single, centralized platform, it combines budgeting, forecasting, BI reporting, and intuitive modeling technologies, empowering users to create and schedule applications that are tailored to the operations of their firm.
       \\ \hline

  \end{tabular}
}
\end{center}
\end{table}

\subsubsection{Simple online tools}
\paragraph{\textbf{Tableau}} Tableau Public is a free platform for exploring, creating, and publicly sharing data visualizations online \cite{kennedy2016data}.



\paragraph{\textbf{Canva} }Unlike other charting tools, Canva is quick and easy. There's no learning curve - you'll have a beautiful graph or chart in minutes, turning raw data into something visual and easy to understand \cite{ahn2019data}.

\paragraph{\textbf{Plot.ly} } Plotly is an open-source visualization library for data visualization and analysis. It provides many products including Dash, Chart Studio,, a Python framework, R, and recently JULIA for building fast, easy and powerful analytical applications.  It gives the hand to draw several types of the graph such as 3D graphs, histograms..., easy to use and handle, totally free, and very interactive and flexible.



\paragraph{\textbf{PowerBI} }Is a data analysis solution from Microsoft. It allows for the creation of personalized and interactive data visualizations with a simple interface enough for end users to create their reports and dashboards \cite{viorel2019analysis}.



\paragraph{\textbf{Chartblocks} } This is an online charting software. It helps to create basic charts quite quickly and to import more data from different external sources \cite{fahad2018big}.

\paragraph{\textbf{Periscope Data} }Is an effective platform for data analysis. It may compile all of the company's data and produce reports. With this tool, we can quickly transform our data into a report or graph that is simple to interpret. \cite{pattanaik2021data}.

For data consumers who frequently need data, Periscope Data enables analysts to transform their SQL searches into interactive dashboards, charts, and reports. The ground-breaking data warehouse technology from Periscope Data links to your databases in a flash to provide extraordinarily speedy, low-cost query processing. Workflow hurdles are removed, and data literacy is promoted throughout your organization, thanks to unlimited users and no query limits. \cite{richardson2020magic}.

\paragraph{\textbf{Holistics} }Is an intelligent data reporting and business intelligence platform that enables us to resolve our data-related inquiries and problems without the need for technical support. For business and data teams, it eliminates the aggravation of request queues. Holistics allows business users to access their data without writing SQL (Structured Query Language) or interfering with data teams. Data teams can create and manage a set of business KPIs using Holistics \cite{khan2021data}.

\paragraph{\textbf{Cluvio} }It is a cloud-based analytics and BI tool that enables businesses to use a dashboard to examine data. The solution enables query execution, results filtering, and data visualization on charts and diagrams. An SQL editor, adaptable R scripts, push notifications, customer dashboard sharing, and more are some of its primary features \cite{srivastava2022review}.

\paragraph{\textbf{Klipfolio} }is a cloud-based tool for building and sharing real-time dashboards for use on mobile, TV, and online browsers. \cite{amer2019tableau}.

\paragraph{\textbf{Clicdata} }ClicData is a 100\% cloud platform. It offers a more modern vision of the software by offering in particular the possibility of importing data regardless of their format and cross-referencing information from different tools \cite{viljanen2020improving}.

\paragraph{\textbf{Qlik Sense} }An effective visualization presents the relationships between many values and allows you to analyze data at a glance. Qlik Sense offers a wide range of visualizations and charts. Each chart excels at visualizing data in various ways for different purposes. Charts should be selected based on the data you want to see in them \cite{vashisht2020integrating}.

\paragraph{\textbf{Chartio} }With the help of the cloud-based BI and analytics tool Chartio, users can quickly evaluate data from business apps and visualize it using a variety of customisable charts. Due to its straightforward Interactive and SQL modes, Chartio is appropriate for both professionals and those with no prior technical knowledge \cite{sedrakyan2020linking}.

\paragraph{\textbf{DataWrapper}}Another excellent tool for data visualization is Datawrapper. With DataWrapper, you can simply generate charts, tables, and maps that are readable on any device.  As a non-commercial platform, Datawrapper is best suited for schools and small businesses that require simple data visualization tools \cite{islam2019overview}.



\paragraph{\textbf{Venngage}}Venngage is a web application for creating a range of data visualizations including infographics, posters, reports, and promotions \cite{usova2021teaching}.

\paragraph{\textbf{Piktochart}} Piktochart is a web-based graphic design tool and infographic maker. We can create bar charts, maps, line graphs, scatter plots, and more \cite{peddoju2020evaluation}. It can also synchronize with Google Spreadsheet or SurveyMonkey to retrieve data and thus create interactive graphs or tables.

\paragraph{\textbf{Infogram}}Infogram is a web-based data visualization and infographics platform. It operates as a data visualization tool to make data easy to understand, discover unknown facts/outliers/trends, visualize relationship patterns, and ask better questions.


\paragraph{\textbf{Raw}}RAWGraphs is an open-source data visualization platform designed to make it simple for anyone to visualize complex data. It tries to fill the gap between vector graphics editors like Adobe Illustrator, Inkscape, and Sketch and spreadsheet programs like Microsoft Excel, Apple Numbers, and OpenRefine  \cite{mauri2017rawgraphs}.

\begin{figure}[ht!]
\centerline{\includegraphics[width=5.5cm]{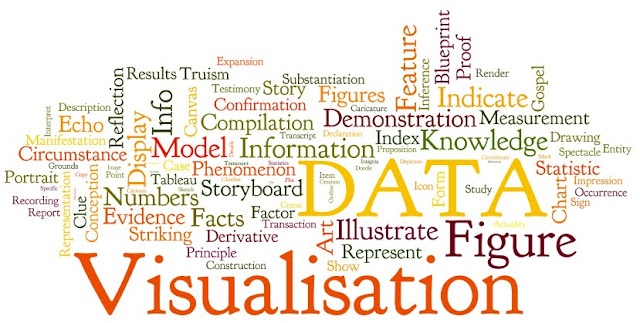}}
\caption{DataViz with Wordle.}\label{fig:wordle}
\end{figure}

\paragraph{\textbf{Wordle}}Wordle is a tool for altering \textit{word clouds} \cite{viegas2009participatory} that include participation from the Las Vegas 2009 Olympics.
The fundamental advantage of Wordle is that it enables neighborhood-preserving editing, which retains words in predictable and nearby areas both during and after editing.  An illustrative example made with wordle for the DataViz context can be presented in figure \ref{fig:wordle}.

\paragraph{\textbf{Easel.ly}}People may simply visualize information on Easel.ly by quickly creating infographics and data visualizations. No prior design experience is necessary. A tool for designing infographics called Easelly can turn any visualized content into any kind of information. It offers a variety of templates, themes, and objects so users may alter specific details in their projects \cite{weiner2021infographics}.

\subsection{Tools accessible to experienced users}
\subsubsection{JavaScript libraries}
JavaScript libraries and frameworks facilitate the development of websites and applications with a wide range of features and functionality - all thanks to the dynamic, flexible and attractive characteristics of JavaScript. According to a 2020 StackOverflow survey, JavaScript remains the most widely used programming language (for the 8th grade), with 67.7\% of respondents using it.
Table \ref{tbl:js} including the top 15 libraries dedicated to visualization will be able to describe them better.

 \begin{table}[!ht]
 \centering
  \scriptsize{
   \caption{\label{tbl:js}Top 15 JavaScript Visualization Libraries.}
  \begin{tabular}{lp{8.5cm}}
      \hline
    \bf JavaScript Library & \bf Description  \\
    \hline\hline
    D3JS &
       Best for document modification driven by data.       \\
    Charts.js     &
Best for plotting logarithmic, date, time, or custom scales on sparse and complex datasets.
        \\
     FusionCharts&
    For needs in data visualization and charting for web and enterprise applications, it works best.
       \\
     Taucharts    &
      Ideal for groups creating intricate data displays.
       \\
    Two.js  &
  Best for a 2-D shape rendering open-source library.
        \\
    Pts.js
     &
     Most effective for assembling items as you view them as points with a simple level of abstraction.
       \\
         Raphael.js
     &
   Best for quickly and efficiently producing intricate drawings and graphics.
       \\
    Anime.js&
 Best for producing potent user interface animation that is compatible with all of the main modern browsers.
       \\
ReCharts
     &
Best for teams wanting to build charts for web applications based on React.
       \\
    TradingVue.js
     &
   Best for creating complex charts, especially for stock and Forex trading applications on the web.
        \\
HighCharts
     &
 Best suited for teams needing a large library of charts to serve web and mobile platforms.
       \\
ChartKick
     &
    Best for producing simple charts using various libraries for programming languages, like Python, Ruby, JS, etc.
      \\
    Pixi.js
     &
   Best for groups searching for JavaScript libraries to produce HTML5-based digital content.
       \\    Three.js
     &
      Best for creating 3-D visuals for applications on the web.
           \\
     ZDog
     &
    The creation and rendering of 3-D pictures for canvas and SVG are not best for open-sourced software.
      \\ \hline
  \end{tabular}
}
\end{table}

\subsubsection{Dashboard builders}
\paragraph{\textbf{Google Data Studio}}It is an online application for data visualization that assists users in transforming data into educative reports and engaging dashboards. By producing engaging reports like this one, it is a powerful tool that enables marketers and business owners to use their data efficiently \cite{serik2021optimal}. In essence, Google Data Studio is a condensed version of software for data visualization, such as Tableau and Clickview. Data Studio is not a data source, in contrast to tools like Google Analytics or HubSpot. It gathers data from several sources, does analysis on it, and then enables you to produce interactive reports, charts, and dashboards rather than collecting the data.
\paragraph{\textbf{Toucan Toco}} Is a cloud-based data visualization tool. Intended for non-technical business executives, the objective of this highly configurable data visualization solution is to provide essential information and data for decision-making \cite{arruabarrena2017expert}. Toucan Toco also develops APIs that allow it to integrate with other IT solutions, such as Cognos Analytics and Salesforce. To retrieve the data to be used and then displayed, this advanced reporting tool is thus able to connect to more than a hundred applications: Excel, Google Analytics, and Microsoft SQL Server.
\paragraph{\textbf{Data Hero}} DataHero is a cloud computing BI software platform specializing in data visualization and dashboards. It is the fastest and easiest way to get insights from data. It offers the possibilty to create charts, reports, and dashboards from business data \cite{pedersen2021data}.
\paragraph{\textbf{Looker}} Is a cloud platform dedicated to data visualization. As a BI platform, Looker integrates several features, including many options dedicated to data visualization including the use of reporting tools, the creation of dashboards, and multi-cloud storage. Coders can use the LookML language to program visualization parameters.
\section{Data Visualization challenges}
An interpretation is necessary before using quantitative data. Data visualizations combine the meaning of unprocessed data into meaningful conclusions. When designers put eye-catching visuals ahead of accuracy, the result is misleading visualizations. In order to convey data in an ethical manner, designers need to steer clear of common data visualization errors.
\subsection{Algorithms and data inputs are susceptible to human error}
Because human inputs are fallible, data visualization can only be as good as the people who provide it. Professionals who are unaware of the variations in applications may employ specific algorithms that emphasize specific information while ignoring other information. They might use a specific technique as a one-size-fits-all method for visualizing data, which can result in concepts being misrepresented. Analysts must take into account what makes each use case distinct and employ a system that can help them achieve their particular objectives in order to minimize human error. Additionally, the use of machine learning and artificial intelligence can aid in lowering the requirement for human factors.
\subsection{Data oversimplification}
To make vast amounts of information easier to understand for viewers, visualizations condense them into simple graphs, scatter plots, and other visual aids. Because of this, some professionals have a tendency to oversimplify. If they concentrate too much on the aesthetic appeal, they might overlook important details. As a result, viewers may draw incorrect inferences and conclusions from the imagery. In the end, this may result in poor decisions, which could be detrimental to businesses. By enlisting the aid of data analytics consulting firms, one can guarantee accurate representation of data while lowering the possibility of oversimplification.
\subsection{Reliance on visualization is inevitable}
Customers are depending more and more on data visualizations to understand their information. They make snap judgments based on visuals and aesthetics. It's a simple and efficient method of learning, and even as technology develops quickly, it should continue to be applicable. But in order for businesses to remain competitive, the trend of consumers depending too much on visualization forces them to employ analytical tools.
\subsection{Data overload}
When working with vast and complicated datasets, researchers frequently face the difficulty of data overload. This may lead to misunderstandings, annoyance, and mistakes in your analysis. How do you make sense of your data and prevent yourself from being overloaded with information? Here are some pointers to assist you in controlling the amount of data you analyze. In data visualization, information overload is a prevalent issue. When designers incorporate an increasing number of datasets into the display, data overload occurs. This makes the visualization difficult to understand as well as difficult to construct \cite{rozic2023artificial}. Lack of attention and prioritizing is a common cause of information overload. You should take a step back and assess each data set if you find that you've fallen victim to data overload.
\section{DataViz' issues}
A good visualization makes the data easy to understand so that viewers can rapidly draw conclusions. The inclusion of excessive information is one of the most frequent errors in data visualization. It is challenging for viewers to come up with takeaways because of this. Similar to how visualizations suffer from overuse of visual effects by designers.
\begin{itemize}
    \item \textit{Why are most visualization designs ineffective?} Because they are created for the incorrect audience, data visualizations frequently lack effectiveness. Poor end-user communication causes dashboards to lose some of their apparent usefulness. The audience for the dashboard must be identified before the data visualization design process can begin.
\item \textit{What mistakes should be avoided in data visualization?} Duplicate data, missed data, unmarked NA values, etc. are examples of common errors. For instance, the three pie chart sectors in this pie chart total up to 193\%, which is illogical. Your final visualizations would be useless if the data contained such inaccuracies.
\item \textit{Does data visualization require coding?} You may quickly construct an interactive data visualization without writing any code. Spreadsheets and reports with a lot of text are insufficient for effectively presenting the data we found. This is why data visualization is necessary to show the data in a form that makes it easier for everyone to understand complex ideas.

\item \textit{Why is misleading data bad?} The audience could not notice the pertinent information if there is too much material offered or if it is irrelevant. It gets increasingly challenging to identify specific trends as there is more data provided at once. The public is frequently misled using sparse but pertinent material by over-informing them.

\item \textit{Can the data be misleading?} Due to the sampling technique employed to collect the data, the results may be deceptive. For instance, the type and size of the sample used in any statistic has a significant bearing on the results. Since many surveys and polls are directed at specific groups of people who give particular answers, the sample sizes tend to be tiny and skewed.

 \begin{table}[!ht]
  \scriptsize{
   \caption{\label{tbl:future}Some educated guesses about what the future may bring negatively.}
  \begin{tabular}{lp{7cm}}
      \hline
    \bf AI's DataViz issue & \bf Description  \\
    \hline\hline
    Potential Job Losses &
       Several artists were upset when an AI-generated image won an art competition last year.     \\
       &With a description in natural language, DALL-E and its offspring can produce realistic-looking artwork and images.  \\
    Resource Websites Tools Redundancy     &
Perhaps as our methods of information acquisition change, AI tools will lessen the significance of webpages.
        \\
        &Perhaps articles, blog posts, brochure websites, and even Wikipedia pages will become much less popular.\\
     A Deluge of Mediocrity&
    This isn't unique to DataViz, but AI may have a very detrimental impact on the caliber of web content.
       \\
       &There will probably be a spike in the quantity of AI-generated content available online as these tools enable faster and simpler \\
       & content production.\\
       &It's possible that online content created by AI will eventually surpass that created by humans.\\
     Fake Data Visualisation    &
      We might observe AI producing a ton of charts in the field of data visualization to further a cause.
       \\
       &We learned how ChatGPT can create fake data in the previous post. Therefore, it is not out of the question that AI could create   \\
       &charts in large quantities using fictitious datasets. Even if a source is included at the bottom of the chart, it's possible that it was \\
       & created entirely by AI to deceive those who were unwilling to look further.\\
    Impact on Design Decisions  &
  Over the last ten or so years, there has been a noticeable decline in attention spans.\\
  & The growing appeal of Tiktok and YouTube shorts, particularly among younger audiences, makes this evident. \\
  &It is unfortunate to say that if this trend keeps up, it will definitely affect how people look for and consume content that is data-focused.
      \\ \hline
  \end{tabular}
}
\end{table}

\end{itemize}

\subsection{How to secure data privacy when sharing visualizations?}
A strong technique to convey patterns, insights, and narratives from large, complicated data sets is through data visualization. However, how can you make sure that unauthorized users cannot access your visualizations and see private or sensitive data?

Individuals and organizations have the right and obligation to manage the collection, use, and sharing of their proprietary or personal data. This is known as data privacy. Data privacy is a business and reputational concern in addition to a legal and ethical one. You run the risk of facing legal repercussions, negative consumer feedback, and reputational harm if you neglect to secure data privacy. Furthermore, you run the danger of losing the confidence of your stakeholders, partners, and data providers, which could have an impact on the availability and quality of your data.

Anonymizing data prior to visualizing it helps protect data privacy. In this procedure, any identifiers such as names, addresses, phone numbers, or email addresses-that could be used to connect data to particular individuals or entities are either hidden or removed. A variety of techniques can be used to anonymize data, including substituting random or pseudonymous values for identifiers, aggregating or summarizing data at a finer level of detail, introducing noise or distortion to value data, choosing or sampling a subset of data, and switching or combining data values between records.

One practical method for guaranteeing data confidentiality is data encryption. This entails converting data into an encoded format that is only readable or accessible by individuals with the proper authorization and a decryption key or password. Data can be encrypted using a variety of techniques, including applying encryption algorithms or software, utilizing secure protocols or platforms, and incorporating encryption functions or libraries into encryption tools. visualization, as well as securing visualization outputs with a password or access control. All of these techniques can aid in preventing unwanted access to sensitive data.
\subsection{AI's impact on data visualisation Work}
Thanks to technology advancements, data visualization once restricted to simple graphs and tables now has a far more sophisticated aspect. As it positions itself as a major catalyst, artificial intelligence reveals new methods for accurately representing and interpreting the vast amount of information at our disposal. The nexus between Artificial Intelligence (AI) and visualization goes beyond a straightforward display; it is revolutionizing our comprehension and utilization of data \cite{rozic2023artificial}.

In terms of theory and consulting, Chat Generative Pre-trained Transformer (ChatGPT) still needed improvement because it was still prone to errors and lacked in-depth knowledge. In spite of this, it's still amazing how quickly and accurately it could respond in writing to my question. On the DataViz coding side was where ChatGPT truly excelled. We can ask ChatGPT to generate the code for a specific chart in multiple languages or libraries, and it will be as simple as that. In order to alter the code's appearance or functionality, we can also ask it to update certain parts of it. Even broken code can be fixed by ChatGPT, which can also tell you what went wrong \cite{kovalerchuk2022integrating,conner2022conceptual}.

What effect will ChatGPT and similar AI tools have on DataViz work in the future, though, if they keep getting better? The way we retrieve information from the internet could already be revolutionized by the ChatGPT interface \cite{lavanya2023comprehensive, kim2023good}. To what extent, then, would AI tools disrupt the data visualization industry? In the following table \ref{tbl:future}, some educated guesses about what could take place in the future because of the AI.

\section{Conclusions and limitations}

With the development of smart technologies that generate astronomical amounts of data, data visualization becomes essential. Indeed, in order to be able to analyze your Big Data and make the best use of it in your business strategy, it is essential to be able to read it and list your business information in visual dashboards.

By classifying, segmenting and scripting data visually, a business can uncover previously inaccessible information at a glance. Data visualization therefore allows any organization to manage its activity more efficiently by adopting a data-driven and agile strategy. If data visualization was important a few years ago, it is crucial today. In the era of Big Data, it makes it possible to make sense of the billions of data that a company can collect every day and which, before this transformation process, are presented in separate lines and are therefore not easily exploitable. Data visualization is a very important task nowadays for data scientists. The main reason for recourse is decision-making. An interpretable, relevant, and innovative visualization can lead to the right decision for a company knowing that this decision could be radical.
Conventional visualization techniques cannot handle the enormous volume, variety, and velocity of data. To do this, several tools have emerged and are constantly evolving.

Therefore, modeling large data is a useful topic right now, among other things. Data modeling is actually a process that gives businesses access to a simple graphical user interface for finding, designing, visualizing, standardizing, and deploying high-quality data assets. Now, a sound data model acts as a guide for creating and implementing databases that use higher-quality data sources to enhance application development and help users make wiser decisions \cite{gahar2023towards}. So, we will be interested in big data modeling systems.
\bibliographystyle{IEEEtran}
\bibliography{references}

\begin{thebibliography}{10}
\providecommand{\url}[1]{#1}
\csname url@samestyle\endcsname
\providecommand{\newblock}{\relax}
\providecommand{\bibinfo}[2]{#2}
\providecommand{\BIBentrySTDinterwordspacing}{\spaceskip=0pt\relax}
\providecommand{\BIBentryALTinterwordstretchfactor}{4}
\providecommand{\BIBentryALTinterwordspacing}{\spaceskip=\fontdimen2\font plus
\BIBentryALTinterwordstretchfactor\fontdimen3\font minus
  \fontdimen4\font\relax}
\providecommand{\BIBforeignlanguage}[2]{{%
\expandafter\ifx\csname l@#1\endcsname\relax
\typeout{** WARNING: IEEEtran.bst: No hyphenation pattern has been}%
\typeout{** loaded for the language `#1'. Using the pattern for}%
\typeout{** the default language instead.}%
\else
\language=\csname l@#1\endcsname
\fi
#2}}
\providecommand{\BIBdecl}{\relax}
\BIBdecl

\bibitem{torphy2020social}
K.~T. Torphy, D.~L. Brandon, A.~J. Daly, K.~A. Frank, C.~Greenhow, S.~Hu, and
  M.~Rehm, ``Social media, education, and digital democratization,''
  \emph{Teachers College Record}, vol. 122, no.~6, pp. 1--7, 2020.

\bibitem{walter2021indigenous}
M.~Walter, R.~Lovett, B.~Maher, B.~Williamson, J.~Prehn, G.~Bodkin-Andrews, and
  V.~Lee, ``Indigenous data sovereignty in the era of big data and open data,''
  \emph{Australian Journal of Social Issues}, vol.~56, no.~2, pp. 143--156,
  2021.

\bibitem{08333}
\BIBentryALTinterwordspacing
R.~Mkhinini~Gahar, A.~Hidri, and M.~Sassi~Hidri, ``Let's predict who will move
  to a new job,'' \emph{CoRR}, vol. abs/2309.08333, 2023. [Online]. Available:
  \url{https://doi.org/10.48550/arXiv.2309.08333}
\BIBentrySTDinterwordspacing

\bibitem{HidriAH22}
\BIBentryALTinterwordspacing
M.~Sassi~Hidri, S.~A. Alsaif, and A.~Hidri, ``Performance analysis of machine
  learning algorithms on networks intrusion detection,'' \emph{Int. J. Comput.
  Appl. Technol.}, vol.~70, no. 3/4, pp. 285--295, 2022. [Online]. Available:
  \url{https://doi.org/10.1504/IJCAT.2022.10056028}
\BIBentrySTDinterwordspacing

\bibitem{AlsaifHH21}
\BIBentryALTinterwordspacing
S.~A. Alsaif, A.~Hidri, and M.~Sassi~Hidri, ``Towards inferring influential
  facebook users,'' \emph{Comput.}, vol.~10, no.~5, p.~62, 2021. [Online].
  Available: \url{https://doi.org/10.3390/computers10050062}
\BIBentrySTDinterwordspacing

\bibitem{AlsaHH22}
\BIBentryALTinterwordspacing
------, ``Stacking-based modelling for improved over-indebtedness
  predictions,'' \emph{Int. J. Comput. Appl. Technol.}, vol.~69, no.~3, pp.
  273--281, 2022. [Online]. Available:
  \url{https://doi.org/10.1504/IJCAT.2022.10052783}
\BIBentrySTDinterwordspacing

\bibitem{ZoghlamiHA16}
\BIBentryALTinterwordspacing
M.~A. Zoghlami, M.~Sassi~Hidri, and R.~Ben~Ayed, ``Sampling-based consensus
  fuzzy clustering on big data,'' in \emph{2016 {IEEE} International Conference
  on Fuzzy Systems, {FUZZ-IEEE} 2016, Vancouver, BC, Canada, July 24-29,
  2016}.\hskip 1em plus 0.5em minus 0.4em\relax {IEEE}, 2016, pp. 1501--1508.
\BIBentrySTDinterwordspacing

\bibitem{GaharAHH18}
\BIBentryALTinterwordspacing
R.~Mkhinini~Gahar, O.~Arfaoui, M.~Sassi~Hidri, and N.~Ben~Hadj{-}Alouane, ``An
  ontology-driven mapreduce framework for association rules mining in massive
  data,'' in \emph{Knowledge-Based and Intelligent Information {\&} Engineering
  Systems: Proceedings of the 22nd International Conference KES-2018, Belgrade,
  Serbia, 3-5 September 2018}, ser. Procedia Computer Science, R.~J. Howlett,
  L.~C. Jain, Z.~Popovic, D.~B. Popovic, S.~N. Vukosavic, C.~Toro, and
  Y.~Hicks, Eds., vol. 126.\hskip 1em plus 0.5em minus 0.4em\relax Elsevier,
  2018, pp. 224--233.
\BIBentrySTDinterwordspacing

\bibitem{manogaran2017big}
G.~Manogaran, D.~Lopez, C.~Thota, K.~M. Abbas, S.~Pyne, and R.~Sundarasekar,
  ``Big data analytics in healthcare internet of things,'' in \emph{Innovative
  healthcare systems for the 21st century}.\hskip 1em plus 0.5em minus
  0.4em\relax Springer, 2017, pp. 263--284.

\bibitem{gahar2019distributed}
R.~Mkhinini~Gahar, O.~Arfaoui, M.~Sassi~Hidri, and N.~Ben Hadj-Alouane, ``A
  distributed approach for high-dimensionality heterogeneous data reduction,''
  \emph{IEEE Access}, vol.~7, pp. 151\,006--151\,022, 2019.

\bibitem{khalid2021big}
Z.~M. Khalid, S.~R. Zeebaree \emph{et~al.}, ``Big data analysis for data
  visualization: A review,'' \emph{International Journal of Science and
  Business}, vol.~5, no.~2, pp. 64--75, 2021.

\bibitem{lensen2020genetic}
A.~Lensen, B.~Xue, and M.~Zhang, ``Genetic programming for evolving a front of
  interpretable models for data visualization,'' \emph{IEEE transactions on
  cybernetics}, vol.~51, no.~11, pp. 5468--5482, 2020.

\bibitem{skender2023investigation}
F.~Skender, V.~Manevska, I.~Hristoski, and N.~Rendevski, ``Investigation of
  dataviz as a big data visualization tool,'' in \emph{International Symposium
  on Intelligent Manufacturing and Service Systems}.\hskip 1em plus 0.5em minus
  0.4em\relax Springer, 2023, pp. 469--478.

\bibitem{ibeh2023dataviz}
A.~M. S. R.~N. Ibeh, ``Dataviz design: A study of intentions,'' in \emph{46 th
  CONFERENCE}, 2023, p. 392.

\bibitem{friendly2008golden}
M.~Friendly, ``The golden age of statistical graphics,'' \emph{Statistical
  Science}, pp. 502--535, 2008.

\bibitem{morita2019reflection}
T.~Morita, ``Reflection on the development of the tool kits of bertin’s
  methods,'' \emph{Cartography and Geographic Information Science}, vol.~46,
  no.~2, pp. 140--151, 2019.

\bibitem{organisciak2022giving}
P.~Organisciak, B.~M. Schmidt, and J.~S. Downie, ``Giving shape to large
  digital libraries through exploratory data analysis,'' \emph{Journal of the
  Association for Information Science and Technology}, vol.~73, no.~2, pp.
  317--332, 2022.

\bibitem{jung2021designing}
S.~Jung, R.~Xiao, O.~Buruk, and J.~Hamari, ``Designing gaming wearables: From
  participatory design to concept creation,'' in \emph{Proceedings of the
  Fifteenth International Conference on Tangible, Embedded, and Embodied
  Interaction}, 2021, pp. 1--14.

\bibitem{zhang2021interactive}
L.~Zhang, B.~Vinodhini, and T.~Maragatham, ``Interactive iot data visualization
  for decision making in business intelligence,'' \emph{Arabian Journal for
  Science and Engineering}, pp. 1--11, 2021.

\bibitem{naerland2021towards}
T.~U. N{\ae}rland and M.~Engebretsen, ``Towards a critical understanding of
  data visualisation in democracy: a deliberative systems approach,''
  \emph{Information, Communication \& Society}, pp. 1--19, 2021.

\bibitem{raineri2021innovation}
P.~Raineri and F.~Molinari, ``Innovation in data visualisation for public
  policy making,'' in \emph{The Data Shake}.\hskip 1em plus 0.5em minus
  0.4em\relax Springer, Cham, 2021, pp. 47--59.

\bibitem{holbrook2019open}
J.~B. Holbrook, ``Open science, open access, and the democratization of
  knowledge,'' \emph{Issues in Science and Technology}, vol.~35, no.~3, pp.
  26--28, 2019.

\bibitem{sarica2019data}
S.~Sarica, B.~Yan, G.~Bulato, P.~Jaipurkar, and J.~Luo, ``Data-driven network
  visualization for innovation and competitive intelligence,'' in
  \emph{Proceedings of the 52nd Hawaii International Conference on System
  Sciences}, 2019.

\bibitem{burnett2021destabilising}
C.~Burnett, G.~Merchant, and I.~Guest, ``Destabilising data: The use of
  creative data visualisation to generate professional dialogue,''
  \emph{British Educational Research Journal}, vol.~47, no.~1, pp. 105--127,
  2021.

\bibitem{deng2005building}
F.~Deng, Z.~Zhang, J.~Zhang, and D.~Zhang, ``Building extraction from multiple
  images and lidar data,'' in \emph{Proceedings of the International conference
  on SAR and Multispectral Image Processing}, vol. 6043, 2005, pp. 515--520.

\bibitem{grant2019pretty}
R.~Grant, ``Pretty persuasion: The advantages of data visualisation,''
  \emph{Impact}, vol. 2019, no.~2, pp. 19--23, 2019.

\bibitem{graessley2019industrial}
S.~Graessley, P.~Suler, T.~Kliestik, and E.~Kicova, ``Industrial big data
  analytics for cognitive internet of things: wireless sensor networks, smart
  computing algorithms, and machine learning techniques,'' \emph{Analysis and
  Metaphysics}, vol.~18, pp. 23--29, 2019.

\bibitem{santolalla2020dataviz}
O.~Santolalla, ``Dataviz,'' in \emph{Rock the Tech Stage}.\hskip 1em plus 0.5em
  minus 0.4em\relax Springer, 2020, pp. 33--49.

\bibitem{oike2019simple}
H.~Oike, Y.~Ogawa, and K.~Oishi, ``Simple and quick visualization of periodical
  data using microsoft excel,'' \emph{Methods and protocols}, vol.~2, no.~4,
  p.~81, 2019.

\bibitem{patel2022data}
N.~Akhtar, N.~Tabassum, A.~Perwej, and Y.~Perwej, ``Data analytics and
  visualization using tableau utilitarian for covid-19 (coronavirus),''
  \emph{{Global Journal of Engineering and Technology Advances}}, vol.~3, pp.
  28--50, 2020.

\bibitem{hiljazi2018developing}
S.~Hiljazi and T.~Curtis, ``Developing an introductory class in business
  intelligence (bi) using ms excel powerpivot.'' \emph{Association Supporting
  Computer Users in Education}, 2018.

\bibitem{dougherty2021hands}
J.~Dougherty and I.~Ilyankou, \emph{Hands-On Data Visualization}.\hskip 1em
  plus 0.5em minus 0.4em\relax " O'Reilly Media, Inc.", 2021.

\bibitem{doshi2014review}
J.~Doshi, A.~Goradia, and D.~Mistry, ``A review of google data visualization
  tools,'' \emph{International Journal of Current Engineering and Technology},
  vol.~4, no.~5, pp. 3134--3138, 2014.

\bibitem{cardona2017evaluacion}
J.~A.~S. Cardona and D.~A.~A. Garcia, ``Evaluaci{\'o}n y selecci{\'o}n de
  herramientas de anal{\'\i}tica visual para su implementaci{\'o}n en una
  instituci{\'o}n de educaci{\'o}n superior,'' \emph{Revista IngEam}, vol.~4,
  no.~1, pp. 1--20, 2017.

\bibitem{orlovskyi2019using}
D.~Orlovskyi, A.~Kopp, and V.~Kondratiev, ``Using dashboards for the business
  processes status analysis,'' 2019.

\bibitem{minelli2013big}
M.~Minelli, M.~Chambers, and A.~Dhiraj, \emph{Big data, big analytics: emerging
  business intelligence and analytic trends for today's businesses}.\hskip 1em
  plus 0.5em minus 0.4em\relax John Wiley \& Sons, 2013, vol. 578.

\bibitem{ur2016big}
M.~H.~u. Rehman, V.~Chang, A.~Batool, and T.~Y. Wah, ``Big data reduction
  framework for value creation in sustainable enterprises,'' \emph{Int. J. Inf.
  Manag.}, vol.~36, no.~6, p. 917–928, dec 2016.

\bibitem{oliver2020kiserleti}
H.~Oliv{\'e}r, ``K{\'\i}s{\'e}rleti gy{\'a}rt{\'a}shoz kapcsol{\'o}d{\'o}
  adatvizualiz{\'a}ci{\'o}s fejleszt{\'e}s a purt{\'a}r rendszerben,''
  \emph{Multidiszciplin{\'a}ris Tudom{\'a}nyok}, vol.~10, no.~4, pp. 238--252,
  2020.

\bibitem{atwood2018using}
T.~P. Atwood and R.~Reznik-Zellen, ``Using the visualization software
  evaluation rubric to explore six freely available visualization
  applications,'' \emph{Journal of eScience Librarianship}, vol.~7, no.~1,
  2018.

\bibitem{wilson2014microsoft}
K.~Wilson, ``Microsoft office 365,'' in \emph{Using office 365}.\hskip 1em plus
  0.5em minus 0.4em\relax Springer, 2014, pp. 1--14.

\bibitem{kennedy2016data}
H.~Kennedy and W.~Allen, ``Data visualisation as an emerging tool for online
  research,'' \emph{The Sage handbook of online research methods}, pp.
  307--326, 2016.

\bibitem{ahn2019data}
M.~Kaufmann, ``Big data management canvas: A reference model for value creation
  from data,'' \emph{Big Data and Cognitive Computing}, vol.~3, no.~1, 2019.

\bibitem{viorel2019analysis}
N.~C. Viorel and N.~Lucia, ``Analysis of information on tourism in the european
  union using the power bi business analysis service.'' \emph{Agricultural
  Management/Lucrari Stiintifice Seria I, Management Agricol}, vol.~21, no.~1,
  2019.

\bibitem{fahad2018big}
S.~A. Fahad and A.~E. Yahya, ``Big data visualization: allotting by {R} and
  python with {GUI} tools,'' in \emph{Proceedings of the IEEE International
  Conference on Smart Computing and Electronic Enterprise}, 2018, pp. 1--8.

\bibitem{pattanaik2021data}
S.~N. Pattanaik and R.~P. Wiegand, ``Data visualization,'' \emph{Handbook of
  Human Factors and Ergonomics}, pp. 893--946, 2021.

\bibitem{richardson2020magic}
J.~Richardson, R.~Sallam, K.~Schlegel, A.~Kronz, and J.~Sun, ``Magic quadrant
  for analytics and business intelligence platforms,'' \emph{Gartner ID
  G00386610}, 2020.

\bibitem{khan2021data}
S.~Khan, ``Data visualization to explore the countries dataset for pattern
  creation,'' \emph{International Journal of Online \& Biomedical Engineering},
  vol.~17, no.~13, 2021.

\bibitem{srivastava2022review}
G.~Srivastava and R.~Venkataraman, ``A review of the state of the art in
  business intelligence software,'' \emph{Enterprise Information Systems},
  vol.~16, no.~1, pp. 1--28, 2022.

\bibitem{amer2019tableau}
A.~M. Amer and M.~M. El-Hadi, ``Tableau big data visualization tool in the
  higher education institutions for sustainable development goals,''
  \emph{International Journal of Computer Science and Mobile Computing}, 2019.

\bibitem{viljanen2020improving}
I.~Viljanen, ``Improving solutions for analytics services in a mid-sized
  insurance company,'' 2020.

\bibitem{vashisht2020integrating}
V.~Vashisht and P.~Dharia, ``Integrating chatbot application with qlik sense
  business intelligence ({BI}) tool using natural language processing
  ({NLP}),'' in \emph{Micro-Electronics and Telecommunication
  Engineering}.\hskip 1em plus 0.5em minus 0.4em\relax Springer, 2020, pp.
  683--692.

\bibitem{sedrakyan2020linking}
G.~Sedrakyan, J.~Malmberg, K.~Verbert, S.~J{\"a}rvel{\"a}, and P.~A. Kirschner,
  ``Linking learning behavior analytics and learning science concepts:
  Designing a learning analytics dashboard for feedback to support learning
  regulation,'' \emph{Computers in Human Behavior}, vol. 107, p. 105512, 2020.

\bibitem{islam2019overview}
M.~Islam and S.~Jin, ``An overview of data visualization,'' in
  \emph{Proceedings of the International Conference on Information Science and
  Communications Technologies (ICISCT)}, 2019, pp. 1--7.

\bibitem{usova2021teaching}
T.~Usova and R.~Laws, ``Teaching a one-credit course on data literacy and data
  visualisation.'' \emph{Journal of Information Literacy}, vol.~15, no.~1,
  2021.

\bibitem{peddoju2020evaluation}
S.~K. Peddoju and H.~Upadhyay, ``Evaluation of iot data visualization tools and
  techniques,'' in \emph{Data visualization}.\hskip 1em plus 0.5em minus
  0.4em\relax Springer, 2020, pp. 115--139.

\bibitem{mauri2017rawgraphs}
M.~Mauri, T.~Elli, G.~Caviglia, G.~Uboldi, and M.~Azzi, ``Rawgraphs: a
  visualisation platform to create open outputs,'' in \emph{Proceedings of the
  12th biannual conference on Italian SIGCHI}, 2017, pp. 1--5.

\bibitem{viegas2009participatory}
F.~B. Viegas, M.~Wattenberg, and J.~Feinberg, ``Participatory visualization
  with wordle,'' \emph{IEEE transactions on visualization and computer
  graphics}, vol.~15, no.~6, pp. 1137--1144, 2009.

\bibitem{weiner2021infographics}
A.~Weiner and K.~Lorber, ``Infographics: A methodology for student research
  presentations and other academic projects,'' in \emph{Proceedings of the
  International Conference on Society for Information Technology \& Teacher
  Education}.\hskip 1em plus 0.5em minus 0.4em\relax Association for the
  Advancement of Computing in Education (AACE), 2021, pp. 649--652.

\bibitem{serik2021optimal}
M.~Serik, G.~Nurbekova, and M.~Mukhambetova, ``Optimal organisation of a big
  data training course: big data processing with bigquery and setting up a
  dataproc hadoop framework,'' \emph{World Trans. on Engng. and Technol. Educ},
  vol.~19, no.~4, pp. 417--422, 2021.

\bibitem{arruabarrena2017expert}
B.~Arruabarrena, ``L’expert en dataviz, un m{\'e}tier en transition,''
  \emph{I2D-Information, donn\'ees documents}, vol.~54, no.~3, pp. 7--8, 2017.

\bibitem{pedersen2021data}
A.~M. Pedersen and C.~Bossen, ``Data work in healthcare: An ethnography of a bi
  unit,'' in \emph{Proceedings of the 8th International Conference on
  Infrastructures in Healthcare}.\hskip 1em plus 0.5em minus 0.4em\relax
  European Society for Socially Embedded Technologies (EUSSET), 2021.

\bibitem{rozic2023artificial}
R.~Rozi{\'c}, R.~Sli{\v{s}}kovi{\'c}, and M.~Rosi{\'c}, ``Artificial
  intelligence for knowledge visualization: An overview,'' in
  \emph{International Conference on Digital Transformation in Education and
  Artificial Intelligence Application}.\hskip 1em plus 0.5em minus 0.4em\relax
  Springer, 2023, pp. 118--131.

\bibitem{kovalerchuk2022integrating}
B.~Kovalerchuk, K.~Nazemi, R.~Andonie, N.~Datia, and E.~Banissi,
  \emph{Integrating Artificial Intelligence and Visualization for Visual
  Knowledge Discovery}.\hskip 1em plus 0.5em minus 0.4em\relax Springer, 2022.

\bibitem{conner2022conceptual}
C.~Conner, J.~Samuel, M.~Garvey, Y.~Samuel, and A.~Kretinin, ``Conceptual
  frameworks for big data visualization: Discussion of models, methods, and
  artificial intelligence for graphical representations of data,'' in
  \emph{Handbook of Research for Big Data}.\hskip 1em plus 0.5em minus
  0.4em\relax Apple Academic Press, 2022, pp. 197--234.

\bibitem{lavanya2023comprehensive}
A.~Lavanya, S.~Sindhuja, L.~Gaurav, and W.~Ali, ``A comprehensive review of
  data visualization tools: Features, strengths, and weaknesses,'' 2023.

\bibitem{kim2023good}
N.~W. Kim, G.~Myers, and B.~Bach, ``How good is chatgpt in giving advice on
  your visualization design?'' \emph{arXiv preprint arXiv:2310.09617}, 2023.

\bibitem{gahar2023towards}
R.~Mkhinini~Gahar, O.~Arfaoui, and M.~Sassi~Hidri, ``Towards big data modeling
  and management systems: From {DBMS} to {BDMS},'' in \emph{Proceedings of the
  {IEEE} International Conference on Advanced Systems and Emergent Technologies
  (IC\_ASET)}, 2023, pp. 1--6.

\end{thebibliography}

\end{document}